\documentclass[twoside,12pt,a4paper]{amsart} 
\usepackage{amsmath,amsfonts,amsthm,eurosym}
\usepackage{graphicx}
\usepackage[colorlinks=true,linkcolor=blue,citecolor=blue,filecolor=blue,urlcolor=blue,pageanchor=true,plainpages=false,linktocpage]{hyperref}
\usepackage[english]{babel}

\numberwithin{equation}{section}

\renewcommand{\epsilon}{\varepsilon}
\renewcommand{\phi}{\varphi}
\renewcommand{\rho}{\varrho}
\renewcommand{\theta}{\vartheta}

\begin{document}
\baselineskip3.6ex


\title{The "emerging" reality from "hidden" spaces}

\author{Richard Pin\v{c}\'ak\\ Institute of Experimental Physics, Slovak Academy of Sciences, 043 53 Kosice, Slovak Republic,
\\E-Mail: pincak@saske.sk \\
\\ Alexander Pigazzini\\ IT-Impresa srl, 20900 Monza, Italy; \\and Mathematical and Physical Science Foundation, 4200 Slagelse, Denmark, \\E-Mail: pigazzinialexander18@gmail.com \\
\\  Saeid Jafari\\ College of Vestsjaelland South, Herrestraede 11, and Mathematical and Physical Science Foundation, 4200 Slagelse, Denmark,\\ E-Mail: jafaripersia@gmail.com \\
\\ Cenap \"Ozel\\ King Abdulaziz University, Department of Mathematics, 21589 Jeddah KSA, \\E-Mail: cenap.ozel@gmail.com}
\renewcommand{\shortauthors}{Pin\v{c}\'ak, Pigazzini, Jafari, \"Ozel}

\date{}

%


\maketitle
%

%

\makeatletter
\def\@makefnmark{}
\makeatother
\newcommand{\myfootnote}[2]{\footnote{\textbf{#1}: #2}}
 \footnote {}


%

\begin{abstract}
The main purpose of this paper is to show and introduce some new interpretative aspects of the concept of "emergent space" as geometric/topological approach in the cosmological field.
We will present some possible applications of this theory, among which the possibility of considering a non-orientable wormhole.
Finally, in \textit{Subsection 2.3}, we give a topological interpretation, using this new approach, to M-Theory and String Theory in $10$ dimensions. Further we present some conclusions which this new interpretation suggests, and also some remarks considering a unifying approach between strings and dark matter. The approach shown in the paper considers that reality, as it appears to us, can be the "emerging" part of a more complex hidden structure.


{\textit {Pacs numbers}}: 11.25.Yb; 11.25.-w; 02.40.Ky; 02.40.-k; 04.50.-h; 95.35.+d
\end{abstract}

\section {Introduction and Preliminaries}
In this paper we want to introduce a new way to understand "emerging spaces" using a geometric/topological approach and to do this we have used PNDP-manifolds.
The PNPD manifolds are a new type of manifolds that can lend themselves to many types of applications. In addition to the mathematical aspect, which concerns the construction and studying of these new types of Einstein warped products manifolds, our physical interpretation of the interactions between positive dimensions and "virtual" negative dimensions, lends these manifolds to the most varied applications and in this paper we deal with some cosmological aspects.
\\
Before coming to present these applications, we recall as preliminary some concepts introduced in \cite{pndp}.
\\
Here we consider an Einstein warped product manifold of special kind, where as a "mathematical tool", for the fiber-manifold ($F$), it uses the derived geometry. Concerning this, we recall that if $A \rightarrow M$ and $B \rightarrow M$ are two transversal submanifolds of codimension $a$ and $b$ respectively, then their intersection $C$ is again a submanifold, of codimension $a+b$. Derived geometry explains how to remove the transversality condition and make sense out of a nontransversal intersection $C$ as a derived smooth manifold of codimension $a+b$. In particular, $dim C = dim M - a - b$, and the latter number can be negative. So the obtained dimensions are not related to the usual geometrical concept of "dimension", but they are "virtual dimensions".
In particular $F$ will be a derived manifold of the form: $\mathbb{R}^d + E$, i.e.  the spaceforms $\mathbb{R}^d$, with orthogonal Cartesian coordinates such that $g_{ij}=-\delta_{ij}$ (where the scalar product with two arbitrary vector fields $\langle V, W \rangle$ is define as: $g_{ij}v^iw^j=-\delta_{ij} v^i w^j= -(v^iw^i)$), by adding a vector bundle of obstructions, $E \rightarrow \mathbb{R}^d$, with $rank(E)=2d$, such that $dim(F)=m=d-2d=-d$ (see \cite{pndp} for more datails about PNDP-manifolds and \cite{Joyce} for Kuranishi neighbourhood and obstruction bundle construction). The most important thing to note is that since each Riemannian geometry operation performed on $F$ is actually performed on the underlying $\mathbb{R}^d$, the authors in \cite{pndp}, considered and defined each operation of Riemannian geometry performed and defined on the underlying $\mathbb{R}^d$ as performed and defined on $F$. 
\\
Now, since $m$ is a virtual negative dimension ($m=-d$), the dimension of the PNDP-manifold ($M, g_M$) will be virtual too ($dim(M)=dim(B)+dim(F)=dim(B)+m=dim(B)-d$, where we have $dim(B)=dim(B')+dim(\widetilde B)$, since $B=B' \times \widetilde B$). Thus $dim(M)$ could be positive, zero or of negative value.
\\
Since $Ric_M=\lambda g_M$, from the definition of PNDP-manifold (see \cite{pndp} \textit{Definition 4}) we have that $\widetilde B$-manifold is also Einstein with the same $\lambda$-constant of the PNDP-manifold, i.e., $Ric_{\widetilde B}=\lambda g_{\widetilde B}$, and when $dim(M)>0$, then $dim(M)=dim(\widetilde B)$; in this way from a speculative/applicative point of view, the authors in \cite{pndp} have considered a special projection that acts as desuspension (the concept of desuspension was introduced in \cite{Margolis}, and it is also considered in \cite{Wolcott}), such that it "projects" $M$ into $\widetilde B$, which the latter has real dimensions, not virtual ones, and this is interpreted in the following way: the manifold $M$ is perceived as the manifold $\widetilde B$ (we remember that both are Einstein, with the same $\lambda$-constant and virtual-$dim(M)$ $=$ real-$dim(\widetilde B$)), the rest of $M$ is hidden, and this in favour of the fact that from the speculative point of view, we consider that the negative virtual dimensions of $F$ act and eliminate each other with the positive dimensions of $B$.
\\
Now we recall the following:
\\
\\
{\bfseries Definition 1:} A warped product manifold $(M, \bar{g})=(B,g)\times_f(F,\ddot{g})$ (where ($B, g$) is the base-manifold, ($F, \ddot{g}$) is the fiber-manifold), with $\bar{g}=g+f^2 \ddot{g}$,  is Einstein if only if:
\\
\\
\numberwithin{equation}{section}
{(1)}
$\bar{Ric}=\lambda \bar{g} \Longleftrightarrow\begin{cases} 
 Ric- \frac{d}{f}\nabla^2 f= \lambda g  \\  \ddot{Ric}=\mu \ddot{g} \\ f \Delta f+(d-1) |\nabla f|^2 + \lambda f^2 =\mu
\end{cases}$
\\
\\
where $\lambda$ and $\mu$ are constants, $d$ is the dimension of $F$, $\nabla ^2f$, $ \Delta f$ and $\nabla f$ are, 
\\
respectively, the Hessian, the Laplacian and the gradient of $f$ for $g$, with $f:(B) \rightarrow \mathbb{R}^+$ a smooth positive function.
\\
Contracting first equation of (1) we get: 
\\
\numberwithin{equation}{section}
{(2)}
$R_Bf^2-f \Delta fd=n f^2 \lambda$ 
\\
where $n$ and $R_B$ is the dimension and the scalar curvature of $B$ respectively, and from third equation, if we consider $d \neq 0$ and $d \neq 1$, we have:
\\
\\
\numberwithin{equation}{section}
{(3)}
$f\Delta fd+d(d-1)|\nabla f|^2+\lambda f^2d=\mu d$
\\
Now from (2) and (3) we obtain:
\\
\\
\numberwithin{equation}{section}
{(4)}
$|\nabla f|^2+[\frac{\lambda (d-n)+R_B}{d(d-1)}]f^2=\frac{\mu}{(d-1)}$,
\\
and for PNDP-manifolds the system (1) becomes:
\\
\\
\numberwithin{equation}{section}
{(1***)}
$\bar{Ric}=\lambda \bar{g} \Longleftrightarrow\begin{cases} 
 Ric'- \frac{d}{f}\tau'^*\nabla'^2 f'= \lambda g' \\ \widetilde \tau^* \widetilde \nabla^2 \widetilde f=0
\\
\widetilde Ric = \lambda \widetilde g \\ \ddot{Ric}=0 \\ f \Delta' f'+(d-1) |\nabla f|^2 + \lambda f^2 =0,
\end{cases}$
\\
\\
(since $Ric$ is the Ricci curvature of $B$, then $Ric=Ric'+\widetilde Ric=\lambda(g'+\widetilde g)+ \frac{d}{f'}\tau'^*\nabla'^2 f'$),
with $Ric'$ is the Ricci tensor of $B'$, $\widetilde Ric$ is the Ricci tensor of $\widetilde B$, $g'$ is the metric tensor referred to $B'$, $\widetilde g$ is the metric tensor referred to $\widetilde B$, $f(x,y)=f'(x)+ \widetilde f(y)$  is the smooth warping function $f:B \rightarrow \mathbb{R}^+$ (where each is a function on its individual manifold, i.e., $f':B'\rightarrow \mathbb{R}^+$ and $\widetilde f:\widetilde B \rightarrow \mathbb{R}^+$), $\nabla^2 f=\tau'^*\nabla'^2f+\widetilde \tau^* \widetilde \nabla^2 \widetilde f$ is the Hessian referred on its individual metric, (where $\tau'^*$ and $\widetilde \tau^*$ are the pullbacks), $\nabla f$ is the gradient (then $|\nabla f|^2= |\nabla' f'|^2 + |\widetilde \nabla \widetilde f|^2$), and $\Delta f=\Delta' f'+ \widetilde \Delta \widetilde f$ is the Laplacian, (since $\widetilde B$ is Einstein, we have $\widetilde \tau^* \widetilde \nabla^2 \widetilde f=0$, then $\widetilde \Delta \widetilde f=0$). 
\\
Therefore for $dim(M)=0$ and $dim(M)<0$, we consider that (2) and (3) become:
\\
\\
\numberwithin{equation}{section}
{(1**)}
$\bar{R}=\lambda \bar n \Longleftrightarrow\begin{cases} 
R'f- \Delta' f'd=n' f \lambda \\ \widetilde \Delta \widetilde f=0 \\ \widetilde R = \lambda \widetilde n \\ \ddot{Ric}=0 \\ f \Delta' f'+(d-1) |\nabla f|^2 + \lambda f^2 =0.
\end{cases}$
\\
\\
where $n'$ and $R'$ are the dimension and the scalar curvature of $B'$ respectively, while for $dim(M)>0$, we must set $d=n'$ and we obtain:
\\
\\
\numberwithin{equation}{section}
{(1*)}
$\bar{R}=\lambda \bar n \Longleftrightarrow\begin{cases} 
R'f- \Delta' f'n'=n' f \lambda \\ \widetilde \Delta \widetilde f=0 \\ \widetilde R = \lambda \widetilde n \\ \ddot{Ric}=0 \\ f \Delta' f'+(n'-1) |\nabla f|^2 + \lambda f^2 =0.
\end{cases}$
\\
\\
\\
Therefore with regard to the projections/desuspensions we remember that:
\\
-if $dim(M)>0$ (i.e. system solutions (1*)) we have the projection:
\\  
$\pi_{(>0)}:$PNDP$\rightarrow (\Pi_{i=(q'+1)}^{\widetilde q}B_i)=\widetilde B$,
\\
- if $dim(M)=0$, (i.e., system solutions (1**)), we have the projection:
\\ 
$\pi_{(=0)}:$PNDP$\rightarrow p$, i.e. a point-like manifold, zero dimension, and
\\
- if $dim(M)<0$, (i.e., system solutions (1**)), we have the projection:
\\
$\pi_{(<0)}:$PNDP$\rightarrow \Sigma^{dim(M)<0}(p)$, with $\Sigma^{dim(M)<0}(p)$, we mean the $(||dim(M)||)$-th desuspension of point, for example, if $dim(M)=-4$ the projection $\pi_{-4}$ will project $M$ into an object which will be given by the fourth desuspension of a point.
\\
\\
Lastly we also recall that referring to a PNDP-manifolds, with negative dimensional fiber, and for not confusing its metric with the metrics of "classics" Einstein warped product manifolds, we denote the Riemannian or pseudo-Riemannian metric of the fiber-manifold with the following notation to indicate that $F$ has negative dimensions: \\ $\ddot{g}=(\Sigma^n_{i=1}(d\psi^i)^2)_{(m)}$, where $m$ is the negative dimension of $F$. Then the general metric form of a PNDP-manifold is:
$\bar g=g-f^2(\Sigma^n_{i=1}(d\psi^i)^2)_{(m)}=(g' + \widetilde g)+(f'+ \widetilde f)^2(\Sigma^n_{i=1}(d\psi^i)^2)_{(m)}$, where $g=g'+\widetilde g$ is the metric of the base-manifold $B$.

\section{PNDP in Cosmology Applications}

In this section we want to give a preliminary view to some possible applications of PNDP-Theory, as geometric/topological approach, in the field of cosmology. Specifically, we will show an application related to the possibility of considering a particular non-orientable wormhole on which we will also hypothesize its behavior, on the hypothesis that an "information" passes through it.
In the last three subsections (2.2, 2.3 and 2.4) we also introduce, in view of this new approach, our future analyses to the fundamental questions of physics and we try to reconsider M-Theory and the String Theory in $10$ dimensions, with some considerations and the possibility of considering point-like universes and also universes with multiple times.

\subsection{PNDP-manifold, non-orientable wormhole and "information" time travel.}

In this subsection we will try to show another interesting application involving wormholes, where the PNDP-manifold is used as an "invisible" connection between the input and output of the wormhole. Here we assume that a PNDP-manifold in which the 2-dimensional base $B$ is a M$\ddot{o}$bius strip.
\\
Specifically the PNDP-manifold, which we call M$\ddot{o}$, will be a point-like manifold as follows: M$\ddot{o}$ $=B \times (\mathbb{R}^2+E)$, where $E$ is a bundle of obstruction with $rank(E)=4$, so, $dim(M \ddot{o})=2-2=0$, zero-dimensional, M$\ddot{o}$ is a point-like manifold (i.e., proj. $\pi_{(=0)}$).
\\
$B=\widetilde B \times  B'$ is a flat 2-dimensional manifold as follow:
\\
$B=(0 \leq y \leq 1) \times (-\infty < x <\infty)$, where is identify $(x,0)$ with $(-x,1)$ for all $x \in \mathbb{R}$, the resulting metric makes the open M$\ddot{o}$bius strip into a (geodesically) complete flat surface and this is the only metric on the M$\ddot{o}$bius strip, up to uniform scaling, that is both flat and complete.
\\
We can see that M$\ddot{o}$ exists, in fact, since $B$ is flat, then M$\ddot{o}$ is flat and $\lambda=0$. 
\\
So, from:
\\
\numberwithin{equation}{section}
{(1**)}
$\bar{R}=\lambda \bar n \Longleftrightarrow\begin{cases} 
R'f- \Delta' f'd=n' f \lambda \\ \widetilde \Delta \widetilde f=0 \\ \widetilde R = \lambda \widetilde n \\ \ddot{Ric}=0 \\ f \Delta' f'+(d-1) |\nabla f|^2 + \lambda f^2 =0.
\end{cases}$
\\
we obtain: 
\\
\numberwithin{equation}{section}
{(1a**)}
$\bar{R}=0 \Longleftrightarrow\begin{cases} 
\Delta' f'd=0\\ \widetilde \Delta \widetilde f=0 \\ \widetilde R = 0 \\ \ddot{Ric}=0 \\ |\nabla f|^2=0,
\end{cases}$
\\
which is solvable for constant $f$, we have that $f'$ and $\widetilde f$ both constants.
\\
Now we proceed with a topological operation on $B$ to cut it along what in topology is called tubular neighborhood:
\\
$T=[0, 1] \times (-\epsilon, \epsilon)$, and we obtain: $B^T=B/T$,
\\
and then we reconsider our PNDP-manifold like this: $M^T\ddot{o}=B^T \times (\mathbb{R}^2+E)$.
\\
This is a M$\ddot{o}$bius strip with "virtual" zero-dimension and in particular it is a cut M$\ddot{o}$bius strip with zero "virtual" dimension, in fact, from our interpretation, we consider the interactions between positive and "virtual" negative dimensions, such that they neutralize each other making $M^T\ddot{o}$ "hidden", "invisible", point-like:
\\
$\pi_{(dim=0)}:M^T\ddot{o} \rightarrow$ point-like manifold (i.e., "invisible" with zero "virtual" dimension).
\\
\\
At this point we consider a topological wormhole $W$, for example something like the Ellis' wormhole (see \cite{Ellis}), which is expressible with the metric: 
\\
$ds^2=-c^2dt^2+d\rho^2+(\rho^2+a^2)[d\theta^2+(sin\theta)^2d\phi^2]$, and we proceed with the gluing operation of the ends of the cut M$\ddot{o}$bius strip, one at the "entrance" and one at the "exit" of the wormhole $W$.
\\
From the topological point of view, the gluing operation is applied considering the subspaces of $W$ and $M^T\ddot{o}$ connected by a homeomorphism.
\\
We consider the subspaces $W_A$, $W_B$, of $W$ and $M^T\ddot{o}_A$, $M^T\ddot{o}_B$ of $M^T\ddot{o}$ (see Fig. 4), through homeomorphisms $u$ and $v$, we get:
\\
$u:M^T\ddot{o}_A \rightarrow W_A$, and $v:M^T\ddot{o}_B \rightarrow W_B$.
\\
The new space obtained by gluing $W$ and $M^T\ddot{o}$, we call $WM^T\ddot{o}$, is the quotient space:
\\
$WM^T\ddot{o}=W \cup (M^T\ddot{o}/ \mathtt{\sim})$,
\\
where $\mathtt{\sim}$ is the equivalence relation on the disjoint union of $W$ and $M^T\ddot{o}$ induced by $u$ which identifies $W_A(p)$ (a point of $W_A$) with $u(W_A(p))$, the same for $v$, with $W_B(q)$ and $v(W_B(q))$, or more precisely:
\\
$W_A(p) \mathtt{\sim} M^T\ddot{o}_A(p) \Longleftrightarrow W_A(p)= M^T\ddot{o}_A(p)$ and $W_B(p) \mathtt{\sim} M^T\ddot{o}_B(p) \Longleftrightarrow W_B(p)= M^T\ddot{o}_B(p)$.
\\
We have, therefore, obtained $WM^T\ddot{o}$ which is a very particular non-orientable wormhole.
\\
In fact, the non-orientable strip glued to the wormhole is a "virtually" point-like PNDP-manifold, considered in our interpretation, invisible.
\\
So, in this approach, we can consider a wormhole, apparently "classic", but which actually "hides" a non-orientable nature that is invisible.
\\
\\
Based on what has been said so far, we could think of a new approach related to time travel. Let's assume that a kind of "information" enters in $WM^T\ddot{o}$ and it connects a point in past, for example 5 years ago, in a "classic" wormhole the information retraces all stages of the past and therefore the events in space-time will recur sequentially for all 5 years up today, in which the "information" recurs at today's instant in which it is about to re-enter the wormhole.
\\
If, instead, the "information" enters in $WM^T\ddot{o}$ (see \textit{Figure 1}), on the way out, the space-time will not retrace the events that have already occurred with the same temporal scan until it returns to today, in fact the information to get back to the point where it entered in $WM^T\ddot{o}$ will take about double so much time, that is about 10 years.
\\
In exit, along the "invisible" cut M$\ddot{o}$bius strip (i.e., the PNPD-manifold), the "information" after 5 years which will be found, at the entrance to the wormhole, but on the opposite side of the strip, it will have to go through the wormhole externally and pass through the PNDP-manifold again, then it takes another 5 years to return to the moment it was about to cross the wormhole, but this time on the correct side of the strip.
\\
This means that, thanks to our $WM^T\ddot{o}$ an "invisible" and imperceptible parallel universe is generated, where the temporal scan will no longer be the same and in which the time passes slower than the starting universe, in about half. Therefore the non-orientable aspect is important in order to alter the temporal scan.
\\
This universe  will be "invisible" because it will develop along the "invisible" PNDP, and then rejoin with the starting one at the moment when the journey began. We consider that the space-time together with "negative" dimensions present in the Bulk, to be twisted and folded, creating the transition from present to past (non-orientable wormhole). The "information" crosses this non-orientable space-time consisting of the "old" space-time and "negative" dimensions and for this reason it sees the events of the past, but what is contained in that "old" space-time does not detect that "information", because the "information" travels within the "interaction" itself and therefore cannot interact with the space-time present in the PNDP point-like (the space-time is already interacting with the "negative" dimensions). Hence the new "invisible" parallel universe, in which "information" travels, cannot be seen and perceived by the original one, and the "information" can only have access by crossing the wormhole.

\begin{figure}[h!]
 \centering
  \includegraphics[width=0.6\textwidth]{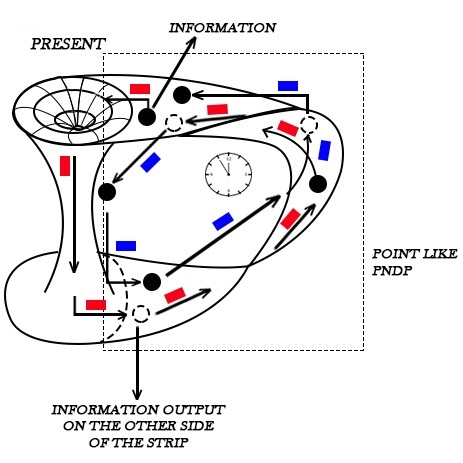}
 \caption{Shown here is the non-orientable wormhole, which shows information that enters the  wormhole and takes a  trip  back in time. The red arrows indicate the  information journey when it is making the first lap,  while in the blue arrows of the second lap are indicated
to return to the correct side of the strip and therefore to be able to return to the wormhole entry instant. Then all space time instants will be retraced from the past to the instant of the journey, but the fact that the information has to make a  double tour, means that in this parallel non-interacting universe, in which information is only a  spectator, the time runs slower than an orientable wormhole. Within the dashed part we indicate that the non-orientable part of the wormhole a  point-like PNDP, i.e. it emerges as a point, zero dimension.}
\end{figure}

The PNDP-point-like manifold allows the creation of non-orientable wormholes that create parallel slow-time universes, invisible, i.e., which cannot be seen and interact, because they develop "within" the interaction between positive and negative dimensions.
\\
This, for instance, is important because it would not violate the principle of causality, because nothing that traces the past could interact with it, as it is not perceptible.

\subsection{PNDP-manifolds and fundamental questions in physics.}

Here we would like to point out about possible applications of PNDP-manifolds in fundamental questions of Physics.
\\
\textit{Example 1.} A very interesting aspect is to reinterpret the example of \textit{Subsection 2.2}, in a string key. Here we want to consider a non-oriented open string (see \cite{Johnson}), as made with the cut M$\ddot{o}$bius strip. In our approach, a string will be a 1-dimensional "object" that "emerges" from a 3-dimensional structure that contains a "virtual" negative dimension. Our goal will be to apply topological operations and glue the ends of the string to a $(1+1)$-wormhole.
\\
In \cite{Guerrea-Ysasi} the authors show how it is possible to consider a $(1+1)$-wormhole in a 2D Minkowski space-time, their model consists on a free massless scalar field that propagates subject to a boundary condition which transfers the field information from one curve $\gamma_{in}$ to another $\gamma_{out}$ via some transfer function $\tau$. The reason for this setup is that a wormhole in one spatial dimension should take particles and information from one point, the entrance, to another, the exit. In the key of our new approach, we can build a PNDP non-oriented string (link patch), and a point-like wormhole, i.e., a PNDP-point-like-manifold, the latter resulting from a $4$-dimensional space containing $2$ "virtual" negative dimensions, the result will be a $(1+1)$-wormhole, as in \cite{Guerrea-Ysasi}, which, due to our interpretation of the interaction between real dimensions and "virtual" negative dimensions, will "emerge" as a point-like space, therefore invisible, or we can also to go further, with an another construction entirely point-like (i.e., wormhole plus link patch both point-like) which is therefore entirely hidden, invisible.
\\
\textit{Example 2.} The construction of a point wormhole could be analyzed to interpret some phenomena known as "quantum fluctuation" (for reference on the subject of quantum fluctuation see for example \cite{Nelson}), and the fact that a pair of "virtual" particles (particle-antiparticle) can suddenly appear and disappear from nowhere, one might think as if at the subatomic level there were such "hidden" wormholes (see \textit{Figure 2}), in which the "interactions" between the positive and the "virtual" negative dimensions, which constitute it (PNDP point-like wormhole), are perceived, upon entry or exit, as "virtual" fluctuating states of energy.

\begin{figure}[h!]
 \centering
  \includegraphics[width=0.9\textwidth]{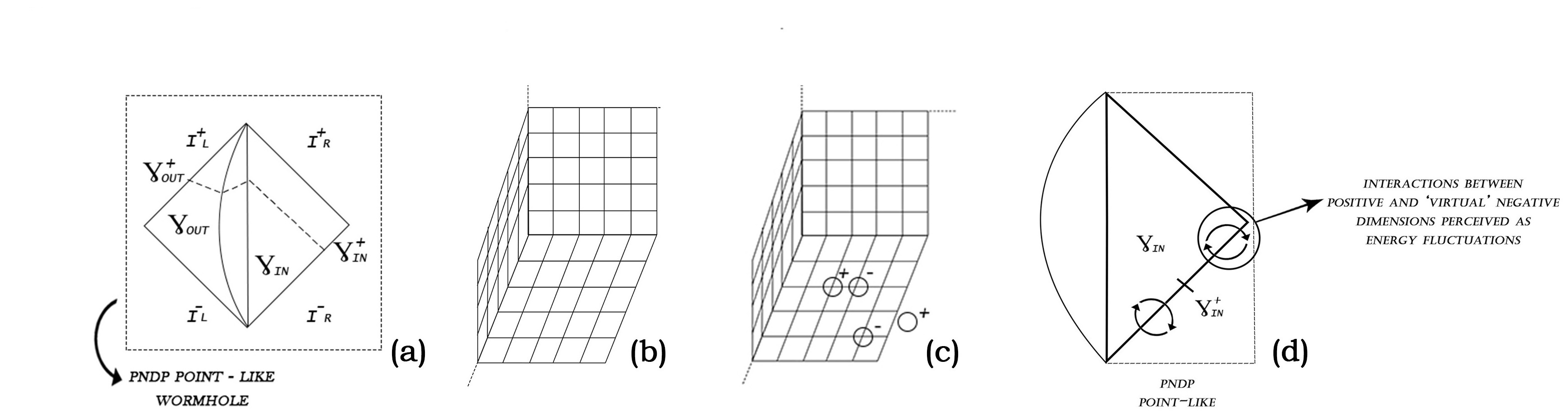}
 \caption{In (a) it was represented the $(1 + 1)$ wormhole introduced and  studied  in \cite{Guerrea-Ysasi} in which we consider it created by a PNDP in a  point-like version.
The curve $\gamma_{IN}$ represents the entrance, while the curve $\gamma_{OUT}$ represents the exit.  In (b) a  vacuum place in space-time is represented.
In (c) pairs of virtual particles are represented, particle-anti particle that annihilate. So (b) and (c) represent the phenomenon known as quantum fluctuation, i.e. fluctuations in the energy of the vacuum that suddenly generate particle-antiparticle pairs that annihilate. Here we assume that this could be generated by (a), i.e. the interactions
between the dimensions that generate the wormhole point like, are perceived, in the entry and exit areas, as virtual energy fluctuations, as in (d).} 
 \end{figure}

The fact of utilizing a non-orientable string is considered as the possible creation of a non-interacting slow-time parallel universe, mentioned in \textit{Subsection 2.1}.
\\
\textit{Example 3.} Another interesting aspect that we would like to point out is also the phenomenon of quantum entanglement (see \cite{Bengtsson}, \cite{Brody} for an introduction). In this approach we could analyze the creation of a PNDP-point-like-manifold (i.e., invisible) every time two particles interact with each other. In the moment of interaction, in fact, this structure could be created which keeps the particles connected wherever they end up, even if they are light years from each other.
\\
\textit{Example 4.} The Dark Matter theme (see \cite{Bertone}, \cite{Bauer}), is also a somewhat controversial theme, and with our approach could give a different interpretation (see \textit{Figure 3}), considering the aspect of "emerging space" as "visible matter" and a "hidden" non-interacting part, such as " Dark Matter ", i.e., that develops "inside" the interaction between the positive and negative dimensions.

\begin{figure}[h!]
 \centering
  \includegraphics[width=0.7\textwidth]{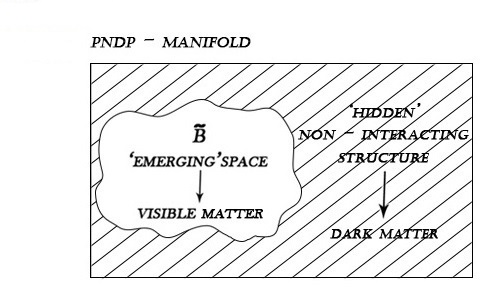}
 \caption{The image shows a PNDP-manifold, in which the emerging part is considered  visible matter, while hidden part of the structure, which we considered non-interacting, represents dark matter.}
 \end{figure}

\textit{Example 5.} Finally, it would be interesting to study the aspects of these PNDP-point-like structures for graphene. In fact in \cite{Graphene WH} the authors argue that superconductor graphene can be produced by molecules organized in point-like structures, and the $(n, -n)$-PNDP manifold can match this requirement, for example we can study the graphene bilayer structure where the graphene wormhole (PNPD-manifold) connects them.

\subsection{PNDP-Strings.}

Strings are "objects" topologically equivalent to a circle (closed string) or a line interval (open string).
\\
In our approach, we consider as strings the PNDP-manifolds whose "emergent" parts are topologically equivalents to open strings.
\\
For instance, we consider a PNDP-manifold with $dim(B) = 2$, where $B =(I_1 \times I_2)$, $I_i$ are the line intervals of $\mathbb{R}$ (i.e., $I \subset \mathbb{R}$), $F=(\mathbb{R}+E)$ with $rank(E)=2$, so $dim(F) = -1$,  and we obtain 
$dim(PNDP) = 2 -1 = 1$, i.e. our PNDP-manifold "will emerge" as a one-dimensional "object" topologically equivalent to an open string, i.e., $\pi_{(=1)}:[(I_1 \times I_2)\times (\mathbb{R}+E)] \rightarrow I_1$.
\\
Therefore, in this approach, to make a string, we need $3$-dimensional space, in which one of these dimensions must be "negative", and only $1$-dimension of these "emerges". So, the string is $(2-1)$-PNDP $=(I_1 \times I_2)\times F$, and, for our interpretation, the "emergent" space, by our projection, will be $I_1$, hence a line interval, topologically equivalent to an open string (see \textit{Figure 4}).
\\
The system (1*) becomes trivial:
\\
\numberwithin{equation}{section}
{(1a*)}
$\bar{R}=0 \Longleftrightarrow\begin{cases} 
\Delta' f'=0\\ \widetilde \Delta \widetilde f=0 \\ \widetilde R = 0 \\ \ddot{Ric}=0 \\ R'=0,
\end{cases}$
\\
and we have that $\widetilde f$ and $f'$ both constant (such that $f=\widetilde f+f'=1$) are solutions for (1a*).
\\
In particular, a string is generated by the interaction that occurs between the dimensions of $(2-1)$-PNDP-manifold, that is two positive dimensions and a "virtual" negative dimension. We say that a string can "vibrate", because in our approach the string is an "emergent" manifold and we want to consider as its intrinsic characteristic, due to the interaction between the "real" and "virtual" dimension, that the "emergent object" will be unstable to these interactions, so this will result in making it "vibrate" in all the dimensions (except the dimensions that interact to "create" the string itself), which has the manifold that contains it. Therefore, this "emergent" $1$-dimensional space, will not be a dimension of the spacetime we are "building", but it will "vibrate" within it.
\\
The Dirichlet and Neumann boundaries will, therefore, be the same as in the classical string theory. The difference here, is that the strings "emerge" from the spatial dimensions, i.e., from the $(2-1)$-PNDP-spatial-manifold, which is a "vibration" of space.
\\
Here we want to show the PNDP approach with respect to M-Theory. Thus the originally newborn spacetime structure can be built starting from: 
\\
(Time $\times$ $(2-1)$-PNDP $\times$ $G_2$)-manifold, a $9$-dimensional manifold that corresponds to the $8$-dimensional brane.
\\
In one approach to M-theory phenomenology, theorists assume that the seven extra dimensions of M-theory are shaped like a $G_2$-manifold (or Joyce-manifolds, see \cite{DJoyce}, \cite{DDJoyce}).
\\
The manifolds with holonomy group $G_2$, are considered by string theorists (see \cite{Witten} and \cite{Vafa}), as compact manifolds and they are used to justify the extra dimensions not perceptible in our daily reality. Now let's build the two PNDP-manifolds:
\\
I) $(2-1)$-PNDP-manifold is the same as above, i.e., $B =(I_1 \times I_2)$, where $I_i$ are the compact line intervals of $\mathbb{R}$, and $F=(\mathbb{R}+E)$ with $rank(E)=2$, so $\widetilde f$ and $f'$ both constant (such that $f=\widetilde f+f'=1$) are solutions for (1b*).
\\
II) Originally, we have the manifold composed of:
\\
(Time $\times$ $(2-1)$-PNDP $\times$ $G_2$), that "emerges" as a $9$-dimensional manifold, and it corresponds to a newborn spacetime with $8$ dimensions, that we call PNDP-Brane, so the "emergent" string, can "vibrate" in $8$ dimension. Thus we have a bosonic string.
\\
III) After this initially process, the $8$-dimensional PNDP-Branes (i.e., the $9$-dimensional (Time $\times$ $(2-1)$-PNDP $\times$$G_2$)-manifold), glued together, in such a way that the $G_2$ compact manifolds, created three new spatial dimensions (something similar to such a gluing procedure is also presented in \cite{Graphene WH} and \cite{Sepehri}).
\\
IV) The obtained manifold will be: 
\\
(6) (Time $\times$ $\mathbb{R}^3$) $\times$ $(2-1)$-PNDP $\times$ $G_2$, that "emerges" as a $12$-dimensional manifold, that corresponds to a spacetime with $11$ dimensions, so the strings can "vibrate" in $11$ spacetime dimensions. In effect a string can "vibrate" in all the dimensions except the dimensions that interact to "create" the string itself. Therefore it can "vibrate" in the $3$ spatial dimensions of Minkowski spacetime, in the $7$ dimensions of $G_2$, plus time-dimension, (the two interacting dimensions, one positive and one "virtual" negative, are excluded because they interact with each other to "create" the string itself). From strong Whitney embedding theorem, this manifold will be smoothly embedded into $28$-dimensional Bulk. 
\\
As a remark, we can state that since the "emergent" $12$-dimensional manifold corresponds to an $11$-dimensional space-time, because the dimension of the string does not affect the value of the space-time dimension, and since a $(1-1)$-PNDP cannot exist, as well as the curvature of space-time does not affect the interaction between positive and negative "virtual" dimensions, we can rewrite the "emerging" PNDP space-time, when $B$ is flat, as: 
\\
(7) Time $ \times (4-1)$-PNDP $\times G_2$, i.e.,  
\\
(8) Time $\times ((\mathbb{R}^3 \times \mathbb{R}) \times_f (\mathbb{R}+E)) \times G_2$, where the interaction $\mathbb{R} \times_f (\mathbb{R}+E)$ represents, for what has been said, the presence of vibrating strings in (8), pointing out once again that, obviously, writing $\mathbb{R} \times_f (\mathbb{R}+ E)$ or $\mathbb{R} \times (\mathbb{R}+E)$ does not change anything, because the space-time metric does not affect the interaction.
Therefore expressing the PNDP space-time as in (6) or in (8) does not change anything. In (6) the presence of the strings is only made explicit. So in the PNDP approach to strings, space-time must have only $2$ interacting dimensions.
\\
So, in summary, in this topological construction, the string can originally "vibrate" in only eight dimensions (point II). So there is only bosonic string, and after gluing the PNDP-branes (point IV), the strings can "vibrate" in eleven dimensions.
\\
\\
The same construction can be repeated assuming another type of PNDP-manifold, i.e., with a toric spatial formation: $(2-1)$-PNDP$=$ $(S^1 \times S^1) \times (\mathbb{R} + E)$, where we have the "emergent" string as $S^1$, which is topologically equivalent to a closed string. We can also consider that originally there were two separate newborn spacetime, one with open string ($A$) and one with closed string ($B$), in the joining process (III), we can have universes composed of only newborn universes ($A$), or only ($B$), but also mixed, in which ($A$) and ($B$), creating a universe with both types of strings.
\\
In this approach, initially there were only many compact "primordial" newborn universes separated from each other, and in each of these there was only one bosonic string. Then, an unimaginable amount of compact "primordial" universes joined together, to give birth to $3$ new spatial dimensions (those of Minkowski). Having created the universe we live in, where strings can now vibrate in 11 dimensions, thus allowing for fermionic strings, we have a $1$ to $1$ correspondence between bosonic strings present in "primordial" universes and fermionic strings of current universe.
\\
In conclusion, we have that our $11$-dimensional spacetime "emerges" as a manifold with $12$ dimensions, but for our string interpretation, we consider that interactions between the dimensions create instability on the "emergent object" and such spacetime dimension will be "lost" by being unstable, i.e., it will "vibrate" within the other dimensions. Therefore the string could be interpreted from the physical point of view, as a "fluctuating fragment of energy" in the dimensions of spacetime, which arises from the interaction between dimensions which we interpret topologically here. Recently another study carried out by L. M. Silverberg and J. W. Eischen (see \cite{Silverberg}) considers the "fragments of energy" as the possible "building bricks" of the universe. From the point of view of our approach, these "floating fragments of energy" can be thought of as "emerging" from interactions that occur between the dimensions of space-time or the Bulk, which for some reason of an intrinsic nature or for some initial situation, some of these dimensions behave in a way that we can mathematically describe them as "virtual" negative which imply that they interact with the other dimensions.
\\
\\
A similar approach can be considered for superstrings in $10$ dimensions, in which to consider again, as in point (III), the branes glued together forming the three new spatial dimensions, such that we will obtain an "emerging" manifold with $11$ dimensions:
\\
Times $\times$ $\mathbb{R}^3$ $\times$ $(2-1)$-PNDP $\times$ CY (where CY is the Calabi-Yau manifold), that corresponds to a spacetime with $10$ dimension where the strings can "vibrate" in $10$ dimensions.  From strong Whitney embedding theorem, this manifold will be smoothly embedded into $26$-dimensional Bulk. 
\\
Also in this case, we can construct the origin brane, by considering the string as the circle ($\widetilde B=S^1$).
\\
\\
{\bfseries Remarks:} The gluing of point (III), can be reconsidered in a version that includes the description of dark matter in this approach to the strings. In fact, during the union of the primordial branes (PNDP-branes), it is necessary to remember that the two interacting dimensions "cancel" each other, "vanishing" in a "point" and thus making "only" the string object "emerge" ". If the branes are all glued to this "point", to get $ 3 $ additional dimensions, we will also get $ 3 $ additional negative "virtual" dimensions, since the glue "point" is the result of the interactions between the two types of dimensions.
\\
From a mathematical point of view this can be described as a sort of iterated PNDP with the special projection of another PNDP, i.e.:
\\
We consider the base-manifold as: $B=\pi_{(dim=3)}:(4-1)$-PNDP $\rightarrow \mathbb{R}^3$, that is, we consider the projection into Minkowski three-dimensional space with vibrating strings inside, in fact the $(4-1)$-PNDP is like in (7) where we implicitly consider vibrating strings: ($\mathbb{R} \times \mathbb{R} \times \mathbb{R} \times \mathbb{R})\times (\mathbb{R}+E)$.
\\
At this point we consider the new PNDP as: $(\pi_{(dim=3)}:(4-1)$-PNDP) $\times (\mathbb{R}^3+E)$, and therefore a space-time mathematically defined as: 
\\
(9) Time $\times [(\pi_{(dim=3)}:(4-1))$-PNDP $\times (\mathbb{R}^3+E)] \times G_2$.
\\
Therefore the string, or rather this "fragment of energy", can "vibrate/fluctuate" even in the negative "virtual" $3$ dimensions and thus generate non-baryon matter.
\\
This "unifying" approach between PNDP-strings and dark matter, also provides that the spatial dimensions of Minkowski, which we perceive in everyday experience, are actually "emerging" as a point-like manifold, therefore externally, if we could observe from the outside our universe, the space-time structure would be equivalent to a compact $ G_2 $-manifold (or CY-manifold) plus time, in which for each point there exist other $3$ positive spatial dimensions (in addition to $7$, or $6$ in CY case) which however are "hidden" due to the interactions with the other $3$ "virtual" negative dimensions; So, it will "appear" as a point-like object, which moves between compact dimensions in the form of a $G_2$-manifold (or CY-manifold). This could indicate an even more extreme theory in which there is a "container universe", in which for each point of its space-time structure (as in the "classical" approach of string theory) there is a compact structure $G_2$ (or CY) plus time, so this "container universe" would have for each point of its structure, the equivalent of a universe like our universe (which would "appear" as a compact structure $G_2$ with a "floating" point-like manifold inside it.
\\
To conclude these observations, we note that in this "unifying" approach, the Minkowski universe that "emerges" as a point-like "object" has no internal structural instability, in fact we have said that the result of the interactions creates an "emergent" object that we say to be "unstable", as a synonym for "vibrating/fluctuating", therefore inside the structure remains stable, only for an external observer it "emerges" as a "vibrating/fluctuating" point-like "object". In view of this, the "container universe" could also be an iterated PNDP and so on. So, this approach that unifies the PNDP-strings with PNDP dark matter, does not prevent the hypothesis that our universe could be a continuous cycle, in which every point of space-time can also be a "point-like" space-time, so our universe is inside itself indefinitely, that is, every point of our space-time is always our own space-time in a point-like version and so on, we could call it a "looped universe".

\begin{figure}[h!]
 \centering
  \includegraphics[width=0.6\textwidth]{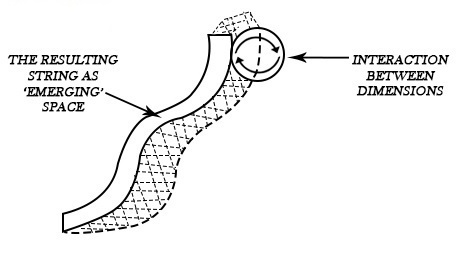}
 \caption{ A  $2-1$- PNDP is shown,ı.e. $(I_1 \times I_2) \times (\mathbb{R} + E)$. From the interaction between the positive and virtual negative dimensions, a line interval emerges, topologically equivalent to  a string.}
 \end{figure}

In \cite{Sepehri}, the authors show that adding a $3$-dimensional manifold to $11$-dimensional spacetime in the Horava–Witten mechanism, all anomalies, from Moffat modified gravity theory (MOG), can be removed and an action without anomaly can be produced. They consider that there are only point-like manifolds with scalars which attach to them. By joining these manifolds, $1$-dimensional manifolds are emerged which gauge fields and gravitons live on them. Then, these manifolds glue to each other and build higher $N$-dimensional manifolds with various orders of gauge fields and curvatures.
\\
What was considered in \cite{Sepehri}, can be considered in our approach, in which the point-like manifolds are PNPD-$D0$-Brane and the strings attached to them are the PNDP-Strings.
\\
In \cite{SepPin} and \cite{GTheory}, the authors introduced a new theory called G-Theory, in which they assumed that the universe was born from an initial state in which two excited strings, one with positive energy and one with negative energy, making their energies flow creating other dimensions and these decaying strings created the so called G0-branes, which by joining together formed the Gp branes. Also in this case our theory can foresee a new approach to G-Theory with an addition of dimensions, including negative ones.

\subsubsection{ T-duality and Mirror Symmetry}

T-Duality and Mirror Symmetry are two fundamental theories in string theory. In general, T-duality relates two theories with different spacetime geometries and in this way it suggests a possible scenario in which the classical notions of geometry break down in a theory of Planck scale physics. 
\\
According to the SYZ conjecture, T-duality is closely related to another duality called Mirror Symmetry, see \cite{SYZ}, and the latter is an important computational tool in string theory. Of course as B. Greene et al. [\cite{Green1}, \cite{Green2}] pointed out that a direct analyze of the strongly coupled theories is hard but by utilizing mirror symmetry they are equivalent to weakly coupled field theories on the mirror CY-space. Here we do not discuss the interesting results concerning physically smooth transitions between Calabi-Yau manifolds having different Hodge numbers (see \cite{Green2}). But it will be an interesting challenge to study these transitions by our point of view which is discussed in this paper. Moreover, the aspect of topology changing transition which relates black holes and elementary particles is also of great interest. The new aspect that our approach introduces lies in the new solutions we can find. In fact, each Calabi-Yau manifold involves different solutions to the string theory equations, resulting in different particles and physical constants.
\\
In our theory/approach, we have a space that "interacts", and generates the string as 1-dimensional "emerging vibrating" space, therefore we have that the vibration of the string depends on the Calabi-Yau manifolds, but also on the "interacting" space. So, in our case, we should not only analyze the vibrations that the string performs inside the Calabi-Yau manifolds (CY), but also the influence that the "interacting" space exerts on the vibrations, i.e., the Riemannian product $(\mathbb{R} \times (\mathbb{R}+E))\times$CY, where $(\mathbb{R} + E)$ is our fiber-manifold ($F$) with "virtual" negative dimension, i.e., a derived-manifold like smooth manifold plus obstruction bundle $E$ (just think of a  Kuranishi neighbourhood $(V,E,s)$, with manifold $V$, obstruction bundle $E \rightarrow V$, and section $s : V \rightarrow E$, then the dimension of the derived manifold is $dim(F)=dim V - rank (E)$. Moreover, since in the definition of fiber-manifold we consider Riemannian flat manifolds, we can extend $F$ not only to the real spaces $\mathbb{R}^d$, but also to the flat-connections). Then, we consider that the string vibrates within CY, but we calculate it in $(\mathbb{R} \times (\mathbb{R}+E))\times$CY, and this so because the Riemannian product does not lead to an increase in the dimension in which the string vibrates. In fact "positive" dimensional space $\mathbb{R}$ and "virtual" negative dimensional space $(\mathbb{R} + E)$, interact and we have mentioned that the "emerging string" could not be considered as a one additional dimension in the Riemannian product, because, in our interpretation, it is "unstable", and also the PNDP-string does not vibrate within the interacting dimensions that "create" it. In this way we are able to study what the new solutions it can lead to, and also to which particles or physical constants they lead. Moreover, we study the moduli space trying to build new mirror manifolds.

\subsection{Point-like Universe and PNDP-Universe with multiple times}

In this subsection, we want to consider, according to this new approach, the possibility of constructing a space-time structure that "emerges" as a point-like "object". This could in fact be easily represented by $(4-4)$-PNDP manifolds, i.e., $B=\widetilde B \times B'=-\mathbb{R}_0^+ \times \mathbb{R} \times \mathbb{R} \times \mathbb{R}$ and $F=\mathbb{R}^4+E$ with $rank(E)=8$ and where by $\mathbb{R}_0^+$ we mean the time. In this way we have described a point-like Minkowski space-time, $(-\mathbb{R}_0^+ \times \mathbb{R} \times \mathbb{R} \times \mathbb{R})\times_k(\mathbb{R}^4+E)$, with special metrics: \\ $ds^2=-cdt^2+ dx^2+dy^2+dz^2+k(\Sigma^n_{i=1}(d\psi^i)^2)_{(4)}$, where the warping function $f=k=$ constant, i.e., in other words, we consider the presence of a multidimensional BULK, also containing those dimensions that we consider "virtual" negative, which are interacting with each other and can give life to universes that we cannot perceive because they are dimensionless (zero dimensional) and hence the possibility of parallel universes that we can call "ghost", being "hidden" and not interacting with our universe. This type of PNDP construction containing a Schwarzschild black hole and a Morris-Thorne wormhole both $(2 + 1)$-dimensional was performed in \cite{pndp}, but considering our systems (1**) and (1*) also other "emerging" universes could be realized, for example AdS.
\\
Another interesting topic, to be considered with this approach, could be that of a universe with multiple times, already considered as examples by Bars et al. in \cite{Bars}. In this case, we could consider a PNDP-time, that is a PNDP-manifold time-like, consisting of two "real" times and a "virtual" negative dimensional time, so that a single time "emerges" from the interactions, the one we perceive, but the interaction "hides" the presence of interacting times that can influence the physics of the system considered.

\section{ Conclusions and Remarks}

We have introduced the concept of "emerging" reality, adapting it to the concepts of Cosmology. It has been shown how the new PNDP structure approach can be used to describe "objects" generated from a hidden multi-dimensional structure.
We dealt with the concept of non-orientable wormhole, with which we suggest a hypothesis of time travel that would not violate the Principle of Causality, introducing the concept of a non-interacting slow-time parallel universe, which deserves further study in our future works.
We tried to propose this approach also in the possibility of reflecting and reviewing some of the fundamental questions of quantum physics, such as quantum fluctuation and entanglement,but also related to dark matter, in which ordinary matter (or "emergent" reality) and dark matter ("hidden" structure) that develops "inside" the interaction between the positive and negative dimensions, would be the parts that make up a single PNDP-manifold.
\\
Finally, we also hypothesized its use to describe entire point-like universes and also universes with multiple times.
\\
In \textit{Subsection 2.3}, we have shown a possible new speculative geometric approach in which M-Theory and the String Theory can be addressed, and the real innovation of our theory lies in considering strings not as "vibrating objects", but as "vibrations/fluctuations of space", or using other words, what we interpret as a string, is a 3-dimensional spatial manifold which contains a "negative" dimension and that makes up the primordial brane (PNDP-Brane). Therefore we interpret the presence of a "negative" dimension as a reason for interaction between the dimensions and because of this, an "unstable" 1-dimensional spatial manifold "emerges". So, we consider these interactions as reasons for instability, and this manifold cannot increase the dimension of the brane within the Riemannian product, because it "vibrates/fluctuates", it is "unstable", otherwise it would create an unstable spacetime. For this reason it is comparable not to a dimension that adds to the PNDP-Brane, but to an energy, arising from interactions, that "fluctuates" through the dimensions of the PNDP-Brane (i.e., CY dimensions and time). We call this manifold string (PNDP-String), so the interpretation leads us to have the equivalence between 1-dimensional "unstable/vibrating" manifold (which is topologically equivalent to the string), and an energy that "emerges" from the interactions of dimensions.
\\
This mathematical construction suggests the possible existence of more universes. In fact, in the composition of the universes, the "primordial" newborn universes ($A$) and ($B$) played the fundamental role; they may have generated three types of universes: only ($A$) universes, which we call ($AA$)-universe, with bosons and fermions from open strings (and also closed, for example by the interaction of two open strings), the ($BB$)-universe, from only ($B$) universes, i.e. with bosons and fermions from only closed strings and ($AB$)-universe, with bosons and fermions from both open and closed strings. This suggests that bosons existed before fermions. The bosons pre-exist our universe. They were already present in the primordial universes, so, for example, the pre-existing Higgs boson gave mass to all the new particles created in the new universe, our universe. In fact, according to the prevailing cosmological theory and our model, the Higgs field permeates all empty space in the universe at all times, even in the initial moments of the Big Bang.
\\
In \textit{Sub-subsection 2.3.1}, we also showed that our approach could lead to new solutions, so, in summary, the novelty we bring is not only in the new string definition, but also in the way it can vibrates.
\\
In \textit{Remarks}, a unifying PNDP approach between strings and dark matter was considered. This could suggest the extreme idea of a "looped universe", as well as motivating the current impossibility of detecting extra dimensions being, in this case, our perceived three-dimensional space to move within additional dimensions.
\\
To conclude, the aspect concerning graphene is also of great interest as this material is the subject of continuous studies.
\\
The aim of our future works will therefore be to delve into each of the topics illustrated, trying to explore the potential that the approach with the PNDP-theory can offer.
\\
\\
\\
\\
{\bfseries \centerline{acknowledgments}}
\textit{The work was partially funded by Slovak Grant Agency for Science VEGA under the grant number VEGA 2/0009/19.}

\bigskip
\bigskip
\bigskip
\bigskip

\end{document}